\documentclass[%
 reprint,
 amsmath,amssymb,
 aps,
 longbibliography,
 superscriptaddress,
]{revtex4-2}

\usepackage{graphicx}
\usepackage{dcolumn}
\usepackage{bm}
\usepackage{subcaption}
\DeclareCaptionFormat{indentlabel}{\hspace{1.5em}#1#2#3}
\usepackage{caption}
\usepackage{amsmath}
\usepackage{braket}
\usepackage{color}
\usepackage[colorlinks=true, linkcolor=blue, citecolor=blue, urlcolor=blue]{hyperref}
\usepackage{chemformula}
\usepackage{placeins}
\usepackage{newtxtext,newtxmath}

\usepackage[a4paper,top=1.5cm,bottom=2cm,left=2.5cm,right=2.5cm]{geometry}

\begin{document}

\preprint{APS/123-QED}

\title{Quantum oscillation fingerprints of altermagnetism in hole-doped \ch{RuO2}}

\author{Yuchi Yang}

\author{Yusheng Hou}
\email{houysh@mail.sysu.edu.cn}

\affiliation{
Guangdong Provincial Key Laboratory of Magnetoelectric Physics and Devices, Institute of Neutron Science and Technology, School of Physics, 
Sun Yat-Sen University, Guangzhou 510275, China
}

\date{}

\begin{abstract}

Altermagnetism, characterized by its ferromagnetism-like spin-splitting band structure and antiferromagnetism-like magnetic order, has garnered considerable attention recently. Although hole doping may promote magnetism in the debated altermagnet candidate \ch{RuO2}, the evolution of its electronic and magnetic properties under hole doping remains poorly understood. Based on first-principles calculations, we employ quantum oscillations to study hole-doped \ch{RuO2}. We find that hole doping can enhance spin splitting and reconstruct the Fermi surface in \ch{RuO2}, which is revealed by the angle-dependent quantum oscillation frequency. By tracking a pair of closed Fermi-surface pockets, we identify a meaningful correlation between the magnetic moment of Ru and a quantum-oscillation-based signature of spin splitting. This correlation follows a quasi-linear trend over a broad doping range, which can be captured by a minimal two-dimensional $d$-wave altermagnetic model. In addition, the hole-doped \ch{RuO2} exhibits a transition from a nonmagnetic to an altermagnetic state via an intermediate state. The quasi-linear correlation through quantum oscillation and the distinct quantum oscillation frequency of the stable altermagnetic state can serve as useful signatures for identifying the altermagnetic state in \ch{RuO2}. Our results provide a comprehensive framework for understanding hole-doped \ch{RuO2}, offering new insights into altermagnetic transitions and their identification.

\end{abstract}

\maketitle

\renewcommand{\theequation}{\arabic{equation}}
\newcommand{\figsubref}[2]{\hyperref[#1]{\ref*{#1}(#2)}}


\section{Introduction}

Collinearly ordered magnetic states used to be thought to fall into two categories, i.e., ferromagnetic (FM) and antiferromagnetic (AFM) orders, with and without spin splitting in electronic bands of the former and latter, respectively. However, this traditional category of collinear magnetism was updated recently by the discovery of altermagnets~\cite{review1,review2,review3,review4}. In an altermagnet, adjacent magnetic sublattices possess oppositely collinear magnetic moments, staggering akin to antiferromagnets. Furthermore, an altermagnet also has spin splitting, similar to ferromagnets in reciprocal space. This particular characteristic makes altermagnets suitable to combine the advantages of both ferromagnets and antiferromagnets. On the one hand, the spin splitting gives rise to many intriguing transport phenomena, such as the anomalous Hall effect~\cite{halleffect1,halleffect2,halleffect3} and tilted spin current~\cite{Tilted-spin-current}. On the other hand, the fully compensated magnetic sublattices in altermagnets preserve zero net magnetization and magnetic invisibility, desirable for ultrafast operations~\cite{Ultrafast_Laser}. Hence, altermagnets uniquely integrate ferromagnet-like spin-polarized transport with AFM robustness, making them highly promising for next-generation spintronic applications~\cite{Antiferromagnetic——spintronics,skyrmion-hall-effect-in-altermagnet}.

As a metallic $d$-wave altermagnet candidate predicted to host large spin splitting, \ch{RuO2} has sparked considerable debate~\cite{Exploration-of-Altermagnetism-in-RuO2}. For decades, this metal has been widely regarded as a Pauli paramagnet on the basis of magnetic susceptibility, heat capacity and electrical resistivity~\cite{MS1,MS2,resistivity,thermal_capacity}. Recently, both theoretical and experimental studies, including polarized neutron diffraction, resonant X-ray scattering, angle-resolved photoemission spectra and so on, have been interpreted as evidence for AFM order in \ch{RuO2}~\cite{AFM1,AFM2,AFM3,AFM4,Observation-of-Spin-Splitting-Torque-in-RuO2}. However, subsequent experiments have cast doubt on the presence of magnetic order in \ch{RuO2}, such as muon spin rotation and relaxation measurements, X-ray diffraction, X-ray linear dichroism and additional angle-resolved photoemission spectra~\cite{NM1,NM2,NM3,NM4,NM5,NM6}. This apparent inconsistency indicates that the magnetism of \ch{RuO2}, if present, may be fragile and highly sensitive to \ch{RuO2} samples. Notably, recent studies indicate that hole doping and epitaxial strain can promote and stabilize altermagnetism in \ch{RuO2}~\cite{Fragility_of_the_magnetic_order,Fragile2}, suggesting a possible origin for different experimental observations.

Given the essential effect of hole doping and strain on the magnetism of \ch{RuO2}, it is of importance to make clear how its electronic and magnetic properties evolve under hole doping. In particular, resolving how spin splitting and Fermi-surface topology change with hole doping is key to identifying the magnetism of \ch{RuO2}. As quantum oscillation provides a direct and detailed probe of the Fermi surface, it is ideally suited to distinguish altermagnetic (AM) and nonmagnetic (NM) states~\cite{Diagnosing_ATM_through_QO,quantum-oscillations}. Previous reports of \ch{RuO2} showed that the experimentally observed quantum oscillation frequency matches better with the frequency calculated in the absence of magnetism~\cite{FS-of-RuO2-by-QO,ab-study-of-QO-inRuO2}. However, the evolution of the quantum oscillation in hole-doped \ch{RuO2} has not yet been explored theoretically and experimentally, leaving the connection between electronic structures, spin splitting, and magnetic states largely unresolved.

In this work, we investigate the quantum oscillation spectrum of hole-doped \ch{RuO2} under a magnetic field rotating from the [$-110$] to [$110$] crystallographic directions, using first-principles calculations. We focus on the evolution of the magnetic moment and a quantum-oscillation-based signature of spin splitting under hole doping in \ch{RuO2}. Our results uncover a robust correlation between the magnetic moment and AM spin splitting through quantum oscillation. This correlation exhibits an overall quasi-linear dependence and can be captured by a minimal two-dimensional (2D) $d$-wave AM model in the small-moment region. In addition, our results indicate that hole-doped \ch{RuO2} experiences a magnetic transition with increasing hole doping, from an NM state to an intermediate state, and eventually to a stable AM state. The quantum oscillation frequency in the stable AM state is distinct from those of the intermediate and NM states. Our results establish a clear connection between quantum oscillations, AM spin splitting, and magnetic states in hole-doped \ch{RuO2}, suggesting that quantum oscillations may provide a useful probe of spin splitting associated with the altermagnetism in hole-doped \ch{RuO2}.

\section{Computational details}

The electronic and magnetic properties of \ch{RuO2} are studied within the framework of density functional theory using the Vienna $Ab$ initio Simulation Package~\cite{VASP}. The exchange-correlation potentials are treated by the generalized gradient approximation with the Perdew-Burke-Ernzerhof functional~\cite{GGA}. We adopt the projector augmented wave method~\cite{PAW1,PAW2} with a plane-wave kinetic energy cutoff of $400\,\mathrm{eV}$. The valence electrons are $4s^24p^64d^75s^1$ for Ru atoms and $2s^22p^4$ for O atoms, resulting in a total of 56 valence electrons per cell in \ch{RuO2} without hole doping. Hole doping is modeled by adjusting the total number of valence electrons in DFT calculations. For \ch{RuO2} under strain, its lattice constants were fixed according to specified strain conditions, while the internal atomic coordinates were relaxed. In such relaxations, we adopt a $\Gamma$-centered $11\times11\times17$ $k$-mesh and energy and force convergence criteria of $10^{-6}\,\mathrm{eV}$ and $0.001\,\mathrm{eV}/\text{Å}$. For the unstrained \ch{RuO2}, the crystal structure as given by Ref.~\cite{Fragility_of_the_magnetic_order} was used without further structural relaxation. The on-site Hubbard $U$ is applied to the $4d$ orbitals of the Ru atoms, with $U=0.4\,\mathrm{eV}$ unless otherwise stated~\cite{LDA+U}. We use a $\Gamma$-centered $25\times25\times35$ $k$-point grid to ensure the accuracy of the calculations of the Fermi surface and quantum oscillations. Quantum oscillations are calculated using the SKEAF code~\cite{SKEAF}, based on the following relation:
\begin{equation}
    F=(\frac{\hbar}{2\pi e})A.
    \label{equ:QO}
\end{equation}
In Eq.~\eqref{equ:QO}, $\hbar$ is the reduced Planck constant, $e$ is the charge of electrons, $F$ is the quantum oscillation frequency, and $A$ is the extremal cross-sectional area of the Fermi surface perpendicular to the magnetic field~\cite{quantum-oscillations}. The magnetic field is rotated from the [$110$] to [$-110$] direction, with the corresponding extremal Fermi-surface cross sections shown as the red and green planes in Fig.~\figsubref{fig:1}{b}.
\nocite{vaspkit}
\nocite{xcrysden}

\section{Results and discussion}

\subsection{Electronic structures and quantum oscillation spectra of NM and AM \ch{RuO2}}

\begin{figure}[htbp]
    \centering
    \includegraphics[width=1\linewidth]{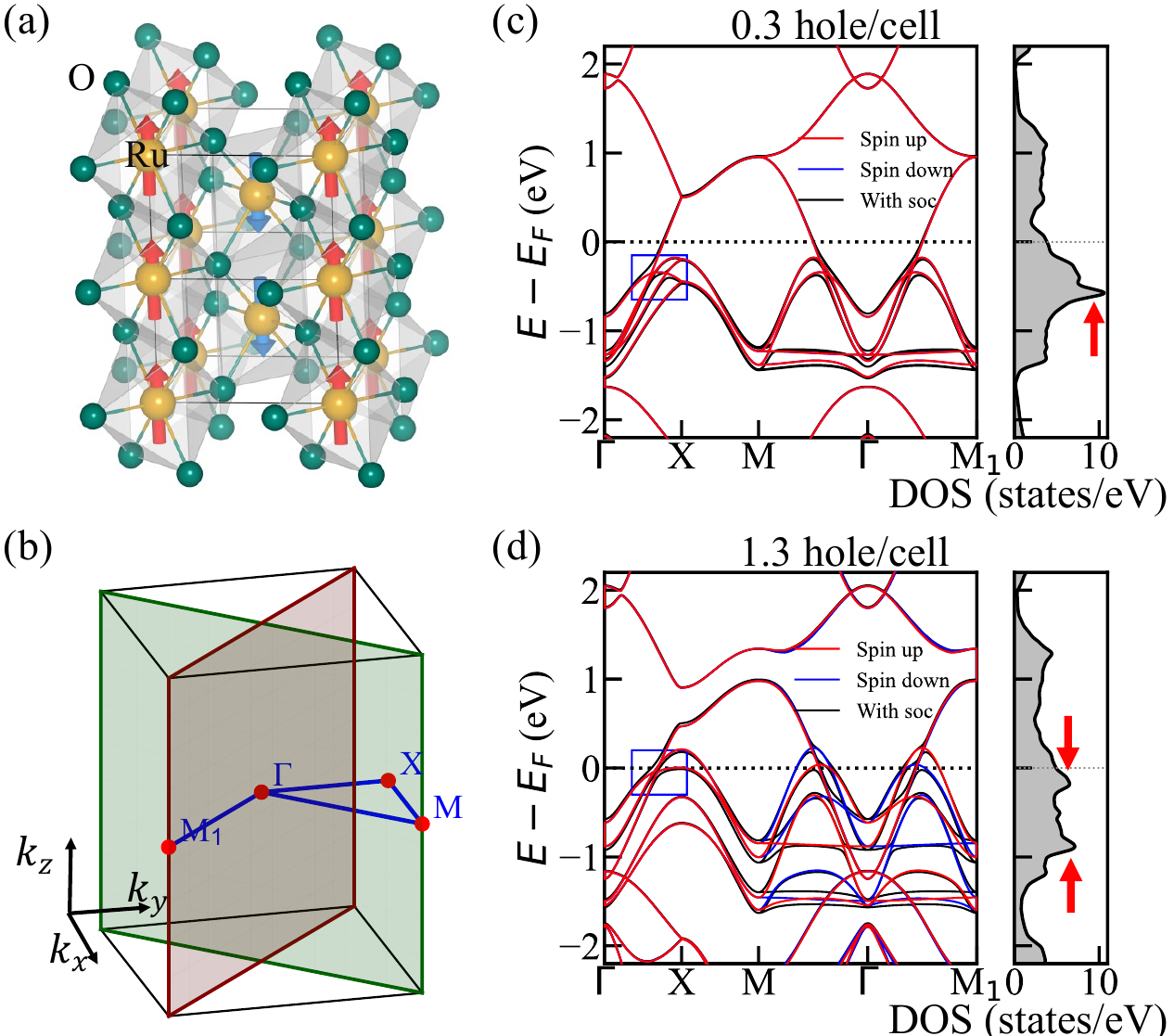}
    \caption{Crystal structure and electronic structures of NM and AM \ch{RuO2}. (a) Crystal structure of \ch{RuO2}, where Ru and O atoms are depicted in yellow and green, respectively. (b) Three-dimensional BZ of bulk \ch{RuO2} and the high-symmetry points and paths. The red and green planes show the [$110$] and [$-110$] crystallographic planes. (c) The left panel shows DFT calculated band structure with and without SOC, and the right panel shows the DOS with including SOC, for NM \ch{RuO2} at 0.3~hole/cell. (d) Same as (c), but for AM \ch{RuO2} at 1.3~hole/cell. The blue rectangles in (c) and (d) highlight a representative SOC-driven gap opening. }
    \label{fig:1}
\end{figure}

\ch{RuO2} crystallizes in the rutile structure with space group $P4_2/mnm$ (No.~136), as shown in Fig.~\figsubref{fig:1}{a}. In the rutile structure, Ru atoms occupy the $2a$ Wyckoff site, and O atoms occupy the $4f$ site with $x=0.30479$, and the lattice parameters are $a=b=4.480\,\text{Å}$, and $c=3.105\,\text{Å}$ \cite{Fragility_of_the_magnetic_order}. The space symmetry of \ch{RuO2} makes it possible to develop a $d$-wave AM state. Despite the expected simple magnetic structure and sizable spin splitting~\cite{review2,AFM1}, the emergence of the AM state in \ch{RuO2} remains under active debate. As has been established in previous DFT calculations~\cite{Fragility_of_the_magnetic_order}, achieving an AM ground state for stoichiometric \ch{RuO2} requires an unphysically large Hubbard interaction for Ru atoms. However, hole doping could promote generating a magnetic order in \ch{RuO2}, potentially driving a transition from NM to AM states~\cite{Fragility_of_the_magnetic_order}. Experimentally, hole doping can be introduced in \ch{RuO2} via intrinsic defects such as Ru vacancies or O excesses, typical defects in this class of materials~\cite{Fragility_of_the_magnetic_order}. In addition, electrostatic gating offers an extrinsic approach for hole doping. By applying a gating electric field, the electron density near the Fermi energy can be tuned effectively, thereby simulating hole doping in \ch{RuO2} as well~\cite{Electrostatic_gating}. Hence, the hole-doped \ch{RuO2} is of theoretical and practical interest for exploring altermagnetism. 

\begin{figure*}[htbp]
    \centering
    \includegraphics[width=1\linewidth]{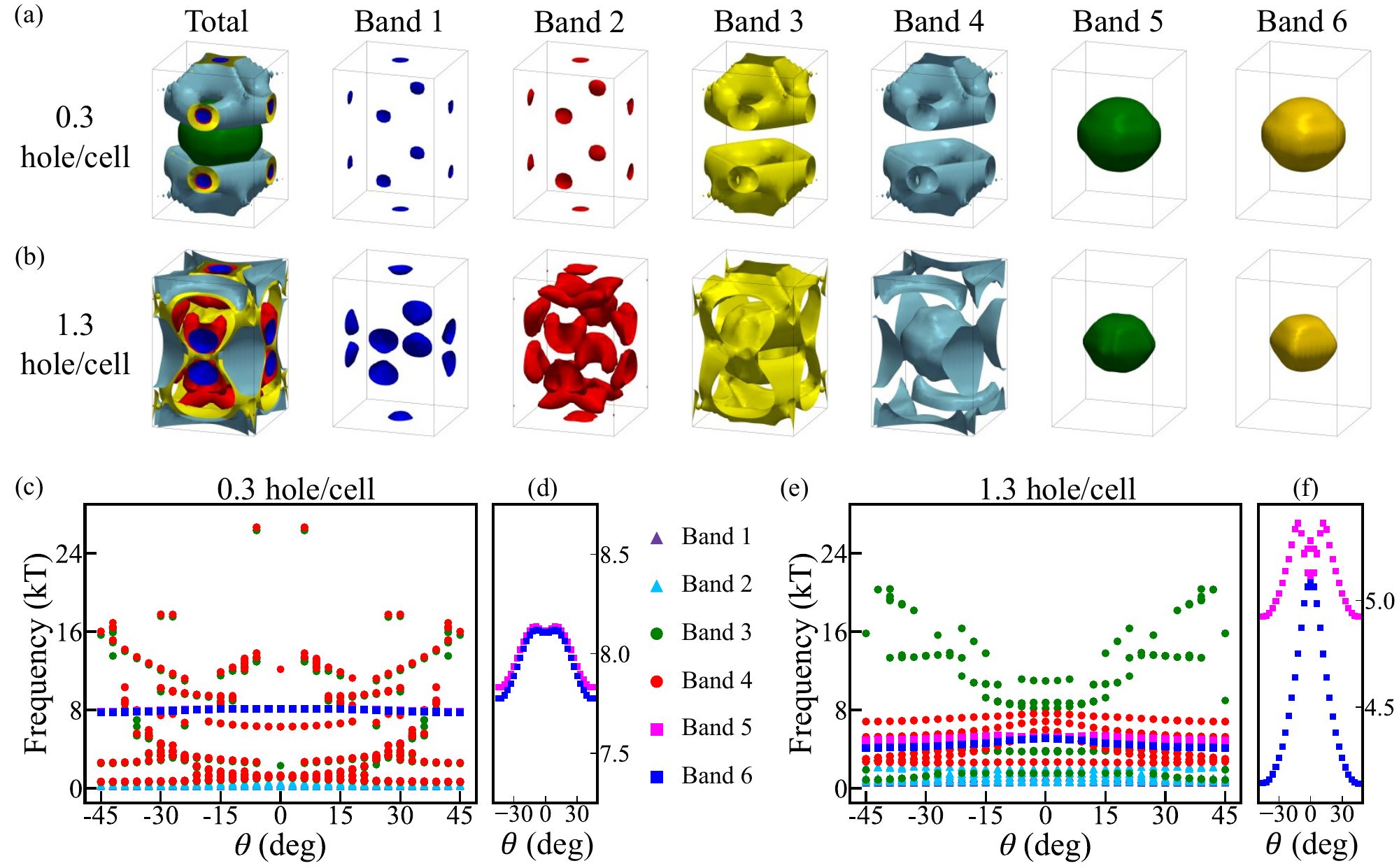}
    \caption{Fermi surfaces and angle-dependent quantum oscillation frequency of hole-doped \ch{RuO2}. (a), (b) Fermi surfaces of NM and AM \ch{RuO2} with the hole doping of 0.3 and 1.3~hole/cell, respectively. Each panel contains the total Fermi surface together with the band-resolved Fermi surfaces for band 1–6. (c), (e) angle-dependent quantum oscillation frequencies corresponding to (a) and (b), respectively. Here, $\theta$ is the angle of deviation from the $k_x$ axis. (d), (f) quantum oscillation frequencies associated with band 5 and 6 at 0.3 and 1.3~hole/cell, respectively.}
    \label{fig:2}
\end{figure*}

To gain insight into the effect of hole doping on the electronic properties of \ch{RuO2}, we choose doping of 0.3 and 1.3 holes per cell as representative cases. At 0.3~hole/cell, the magnetic moment of Ru is almost zero, independent of the initial FM or AM states. This indicates that \ch{RuO2} at this doping level has an NM ground state. In contrast, a staggered magnetic moment of $0.61\,\mu_B$ develops at 1.3~hole/cell, indicating an AM state. In addition, the AM state is energetically favorable, with an energy of $\sim\!16\,\mathrm{meV}$ per cell lower than the FM state. This evolution establishes the transition from an NM to an AM state with increasing hole doping. Fig.~\figsubref{fig:1}{c} and \figsubref{fig:1}{d} give the band structures and densities of states (DOS) of \ch{RuO2} at 0.3 and 1.3~hole/cell, respectively. For the NM \ch{RuO2}, spin-up and spin-down bands are degenerate over the entire Brillouin zone (BZ) in the absence of spin-orbit coupling (SOC). Meanwhile, the DOS below the Fermi energy exhibits a narrow peak, as indicated by the red arrow in Fig.~\figsubref{fig:1}{c}. For the AM \ch{RuO2}, a pronounced spin splitting along the high-symmetry path $M\text{--}\Gamma\text{--}M_1$ is clearly visible as expected. Besides, the DOS in the vicinity of the Fermi energy becomes flatter, and the narrow peak observed in the NM state splits into two lower peaks, as marked by the arrows in Fig.~\figsubref{fig:1}{d}. This split in the DOS is consistent with the modified Stoner picture proposed previously for the magnetic instability~\cite{Fragility_of_the_magnetic_order}. Finally, the bands calculated with and without SOC largely overlap. Only minor modifications are observed, which become more pronounced in the AM state. Along certain high-symmetry paths, SOC also opens gaps at the band crossings calculated without SOC, such as the ones along $\Gamma\text{--}X$ marked by the blue rectangle in Fig.~\figsubref{fig:1}{c} and \figsubref{fig:1}{d}. Overall, SOC has a weak effect on the bands in hole-doped \ch{RuO2}. Despite its relatively weak effects, SOC is included in all subsequent calculations of the hole-doped \ch{RuO2}.

Fig.~\figsubref{fig:2}{a} and \figsubref{fig:2}{b} depict the Fermi surfaces of \ch{RuO2} at doping levels of 0.3 and 1.3~hole/cell, respectively. The Fermi surfaces correlate with six bands which intersect the Fermi energy. For convenience hereafter, these six bands are numbered as band 1-6 and grouped into three pairs (i.e., 1-2, 3-4, and 5-6). For the NM \ch{RuO2} with the doping level of 0.3~hole/cell, the first pair of bands (i.e., band 1-2) forms tiny pockets near the BZ boundary, the second one (i.e., band 3-4) consists of perforated cap-like sheets, and the last pair (i.e., band 5-6) exhibits a simple closed pocket around the $\Gamma$ point. Additionally, we observe that each pair of bands has almost the same geometry in reciprocal space [Fig.~\figsubref{fig:2}{a}]. This is reasonable as the two bands within each pair are actually degenerate in the NM \ch{RuO2}. In the AM \ch{RuO2} at the doping level of 1.3~hole/cell, the first pair (i.e., band 1-2) of bands expands in size with band 2 evolving into a multi-lobed shape, while the second pair (i.e., band 3-4) extends across the BZ and develops an additional closed pocket around the $\Gamma$ point. Notably, the last pair of bands (i.e., band 5-6) preserves a simple closed-pocket geometry in hole-doped \ch{RuO2} even at the doping level of 1.3~hole/cell. Overall, the six bands of the AM \ch{RuO2} display different geometries in reciprocal space. Therefore, the Fermi surfaces of \ch{RuO2} at the doping level of 0.3 and 1.3~hole/cell are clearly different, reflecting different spin splitting in hole-doped \ch{RuO2}. 

In addition to the different Fermi surfaces, the hole-doped \ch{RuO2} with 0.3 and 1.3~hole/cell also exhibits distinct angle-dependent quantum oscillation spectra. For the NM \ch{RuO2} with a hole doping of 0.3~hole/cell, its two frequency branches within each pair of bands basically overlap [Fig.~\figsubref{fig:2}{c}]. By contrast, for the AM \ch{RuO2} with a hole doping of 1.3~hole/cell, its two frequency branches within each pair of bands are clearly separated [Fig.~\figsubref{fig:2}{e}]. Moreover, its frequency becomes smoother and more compact below $10\,\mathrm{kT}$, reflecting the drastic reconstruction of its Fermi surface. Fig.~\figsubref{fig:2}{d} and \figsubref{fig:2}{f} display the quantum oscillation spectra of band 5 and 6 as continuous curves, owing to their simple closed-pocket Fermi surfaces. Given that the contrast between the two Fermi surfaces from each pair of bands reflects spin splitting, the difference in their quantum oscillation frequencies can likewise be used to characterize spin splitting. This connection originates from the fact that quantum oscillation frequencies are determined by the extremal cross-sectional areas of the Fermi surface. Therefore, the separation of the Fermi surface induced by spin splitting naturally manifests itself as a splitting of the quantum oscillation frequencies. So, the quantum oscillation spectra of the hole-doped \ch{RuO2} indicate a nearly vanishing spin splitting at 0.3~hole/cell, and a much stronger one at 1.3~hole/cell, consistent with the above-mentioned differences between their Fermi surfaces. The full evolution of the quantum oscillation spectra over the hole-doping range from 0.0 to 1.5~hole/cell is presented in Fig.~S1 in the Supplementary Material (SM)~\cite{SupplementaryMaterial}. During the evolution, the frequencies reveal a gradual enhancement of the spin splitting with increasing hole doping.
\nocite{minimal_model}

\subsection{Correlation between magnetic moment and spin splitting in hole-doped \ch{RuO2}}

\begin{figure*}[htbp]
    \centering
    \includegraphics[width=1\linewidth]{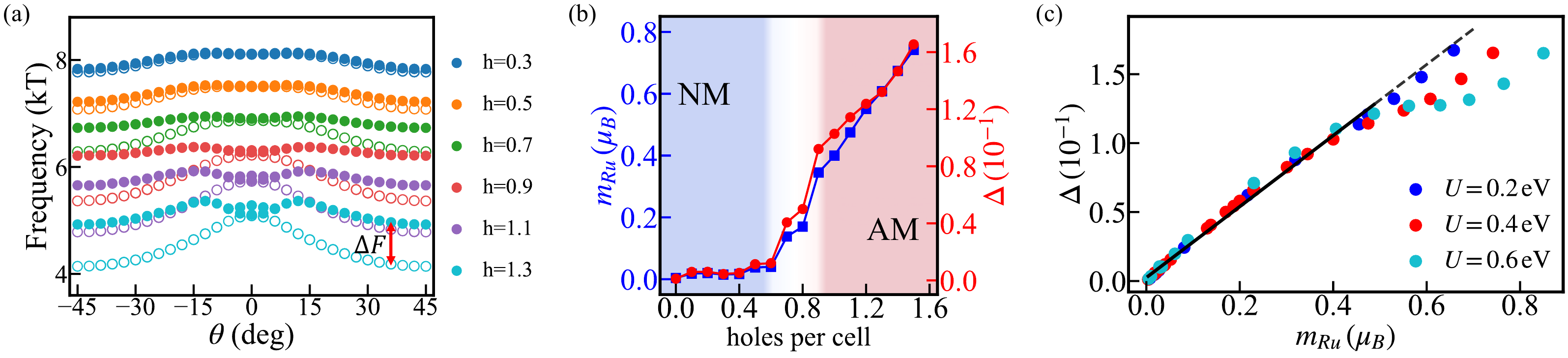}
    \caption{Evolution of the magnetic moment and quantum-oscillation-based signature of spin splitting in \ch{RuO2} under hole doping. (a) Quantum oscillation frequencies of band 5 and 6 with different hole (h) doping. Here, frequencies for band 5 and 6 are depicted in solid and hollow dots, respectively. (b) Hole-doping dependence of $m_{Ru}$ (left axis) and $\Delta$ (right axis). The NM and AM states are indicated by the blue below 0.6~hole/cell, and the red above 0.9~hole/cell, respectively. (c) Dependence of $\Delta$ on $m_{Ru}$ in hole-doped \ch{RuO2} with different on-site Hubbard values. The solid line shows a linear fit in the range $m_{Ru}<0.5\,\mu_B$, while the dashed line represents its extrapolation to large $m_{Ru}$.}
    \label{fig:3}
\end{figure*}

As discussed above, the spin splitting is increasingly pronounced with increasing hole doping, and previous work has also shown an increase in the magnetic moment of Ru over the same doping range~\cite{Fragility_of_the_magnetic_order}. This implies a possible connection between the magnetic moment and spin splitting in hole-doped \ch{RuO2}. For a more thorough analysis, we focus on band 5 and 6 among the six bands. Since their Fermi surfaces are simple closed pockets in the studied hole doping range, their extremal orbits and quantum oscillation frequency branches are well defined in this case. Hence, their quantum oscillation frequency difference, $\Delta F$, can be easily identified. Fig.~\figsubref{fig:3}{a} tracks the doping dependence of $\Delta F$ between band 5 and 6. Upon increasing hole doping, $\Delta F$ becomes obviously larger, and quantum oscillation frequencies shift downward. Moreover, $\Delta F$ is strongly angle-dependent: it becomes larger toward the boundaries of the angular range, while the central region remains nearly overlapping. Such angle dependence indicates the underlying $d$-wave Fermi surface topology in AM \ch{RuO2}. Therefore, the trend of $\Delta F$ clearly reveals a gradually enhanced spin splitting with increasing hole doping in \ch{RuO2}, providing a basis for analyzing the evolution of spin splitting in the following discussion.  

To get a deeper insight into the spin splitting in hole-doped \ch{RuO2}, we introduce a normalized relative difference, $\Delta$, of quantum oscillation frequencies which is defined as follows:
\begin{equation}
    \Delta = \frac{\left\langle \left|F_5 - F_6\right| \right\rangle}{\left\langle (F_5 + F_6)/2 \right\rangle},
    \label{equ:Delta}
\end{equation}
where $F_5$ and $F_6$ are the quantum oscillation frequencies associated with band 5 and 6, respectively, and $\left\langle\cdots\right\rangle$ denotes the average over the full angular range in Fig.~\figsubref{fig:2}{c} and \figsubref{fig:2}{e} with a $3^\circ$ sampling interval. This sampling interval is sufficiently converged, because reducing the interval to $1^\circ$ changes the resulting $\Delta$ by only about $1\%$. A comparison of the results obtained using the $3^\circ$ and $1^\circ$ sampling intervals is presented in Fig.~S3(a) in the SM~\cite{SupplementaryMaterial}. Here, $\Delta$ is used to enable a more meaningful comparison across different doping levels, despite the downward trend in quantum oscillation frequency. Besides, this averaging range provides a more complete characterization of the Fermi-surface splitting, given the $d$-wave symmetry of the Fermi surfaces. As the frequency is proportional to the extremal cross-sectional area of the Fermi surface, Eq.~\eqref{equ:Delta} can be rewritten as follows:
\begin{equation}
    \Delta = \frac{\left| \langle A_5 \rangle - \langle A_6 \rangle \right|}{\left( \langle A_5 \rangle + \langle A_6 \rangle \right)/2},
    \label{equ:Delta2}
\end{equation}
where $A_5$ and $A_6$ are the extremal cross-sectional areas of band 5 and 6, and $\left\langle A\right\rangle$ gives the average extremal cross-sectional area. Thus, Eq.~\eqref{equ:Delta} or \eqref{equ:Delta2} can be interpreted as the normalized average difference of the extremal cross sections, offering a simple yet physically effective method to characterize the spin splitting in hole-doped \ch{RuO2}. 

Fig.~\figsubref{fig:3}{b} presents the evolution of the quantum-oscillation-based signature of spin splitting and magnetic moment of Ru (denoted as $m_{Ru}$) under different hole doping levels. We see that both $\Delta$ and $m_{Ru}$ increase monotonically as hole doping increases. Moreover, their dependence on hole doping displays a highly similar manner, exhibiting synchronous variations and nearly comparable growth. This similarity suggests that the growth of $m_{Ru}$ is closely associated with the enhancement of $\Delta$. To examine this relation in detail, we plot $\Delta$ against $m_{Ru}$ in Fig.~\figsubref{fig:3}{c}. The data obtained with $U=0.4\,\mathrm{eV}$ exhibit an overall monotonic, quasi-linear increasing trend, especially in the small magnetic moment region below around $0.5\,\mu_B$. Because $\Delta$ serves as a quantum-oscillation-based signature for spin splitting in hole-doped \ch{RuO2}, the quasi-linear trend in Fig.~\figsubref{fig:3}{c} reveals a near-linear correlation between $m_{Ru}$ and AM spin splitting in hole-doped \ch{RuO2}. 

To examine the robustness of the above-mentioned quasi-linear correlation in hole-doped \ch{RuO2}, we further investigate several additional cases. These cases include (i) different on-site Hubbard values ($U=0.2\,\mathrm{eV}$ and $0.6\,\mathrm{eV}$), (ii) hydrostatic strain on the entire lattice, and (iii) in-plane epitaxial strain with the out-of-plane $c$ axis fixed. As shown in Fig.~S4 in the SM~\cite{SupplementaryMaterial}, the hole-doping evolutions of $\Delta$ and $m_{Ru}$ in these cases display similar results to the ones as shown in Fig.~\figsubref{fig:3}{b}. In addition, Fig.~\figsubref{fig:3}{c} shows the relation between $\Delta$ and $m_{Ru}$ for different Hubbard $U$ values. The data for different $U$ seem to collapse onto the same line in the small magnetic moment region. A linear fitting below $m_{Ru}=0.5\,\mu_B$, combining all $U$ values, yields $\Delta(10^{-1} )=(2.575\pm0.036)m_{Ru}+(0.025\pm0.008)$ with a coefficient of determination $R^2=0.992$ in Fig.~\figsubref{fig:3}{c}. The $R^2$ value close to unity indicates excellent linearity with a negligible intercept. This fitting establishes a linear correlation between $\Delta$ and $m_{Ru}$ for different $U$ values in the small magnetic moment region, also suggesting a limited effect of Hubbard $U$ on the correlation. As presented in Fig.~S5 in the SM~\cite{SupplementaryMaterial}, the strain-dependent results display similar trends to that in Fig.~\figsubref{fig:3}{c}, thus providing additional support for the robustness of the quasi-linear correlation. Taken together, these results reveal a quasi-linear correlation between the magnetic moment of Ru and AM spin splitting, which is robust in hole-doped \ch{RuO2}.

\begin{figure*}[htbp]
    \centering
    \includegraphics[width=1\linewidth]{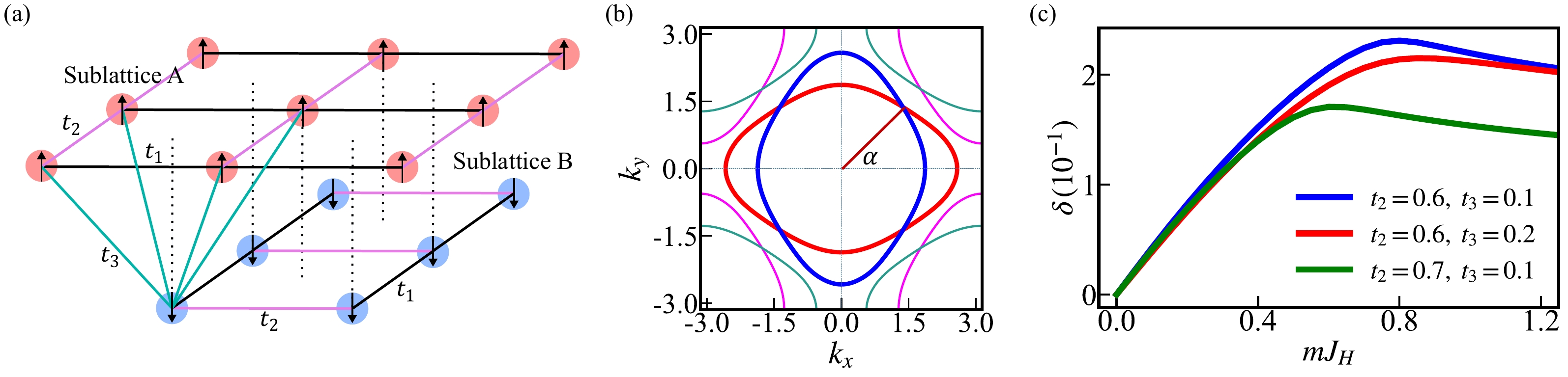}
    \caption{Minimal 2D tight-binding model for the AM state. (a) Schematic of the 2D lattice. (b) 2D Fermi contours at $t_2=0.6$, $t_3=0.2$, and $mJ_H=0.6$. (c) Dependence of $\delta$ on $mJ_H$ for different parameter sets. The quantities $t_2$, $t_3$, and $J_H$ are in units of $t_1$, while the Fermi energy and lattice constant are set to $E_f=0$ and $a=1$ for simplicity. }
    \label{fig:4}
\end{figure*}

An understanding for this quasi-linear correlation can be obtained from a minimal 2D tight-binding model of a $d$-wave altermagnet. The lattice of this 2D model is illustrated in Fig.~\figsubref{fig:4}{a}. Red and blue sites denote sublattice A and B, respectively, and both intralayer and interlayer spacings are set to $a$ for simplicity. Different sublattices occupy different layers and carry opposite magnetic moments, forming a staggered arrangement along the vertical direction. Sublattice B is vertically offset from the centers of the squares formed by sublattice A. In this model, the in-plane hopping $t_1$ and $t_2$ are direction-dependent on each sublattice, deriving from the anisotropic magnetic environment required for $d$-wave altermagnetism. The hopping parameters are also constrained by the symmetry of the $d$-wave altermagnet. In particular, the in-plane hopping along the $x$ direction on sublattice A equals that along the $y$ direction on sublattice B in strength. The inter-sublattice hopping $t_3$ couples nearest-neighbor sites on sublattices A and B. In Fig.~\figsubref{fig:4}{a}, bonds with the same coupling strength are assigned the same color. Overall, this model is designed to capture the essential symmetry features of a $d$-wave altermagnet, rather than the full microscopic details of \ch{RuO2}.

The Hamiltonian of the above-mentioned 2D AM model is in the following form~\cite{Altermagnet_skyrmions:Macroscopic_realization_of_d-wavesymmery}:
\begin{equation}
    \begin{aligned}
    H
    =&H_{kin}+H_{sd}= \\
    &-t_1\sum_{\langle ij\rangle_x,\sigma} c^{\dagger}_{i,A,\sigma}c_{j,A,\sigma}
    -t_2\sum_{\langle ij\rangle_y,\sigma} c^{\dagger}_{i,A,\sigma}c_{j,A,\sigma}\\
    &-t_2\sum_{\langle ij\rangle_x,\sigma} c^{\dagger}_{i,B,\sigma}c_{j,B,\sigma}
    -t_1\sum_{\langle ij\rangle_y,\sigma} c^{\dagger}_{i,B,\sigma}c_{j,B,\sigma}\\
    &-t_3\sum_{\langle ij\rangle_{AB},\sigma}\left(c^{\dagger}_{i,A,\sigma}c_{i,B,\sigma}+\mathrm{H.c.}\right) \\
    &-J_H\sum_{i,\alpha} c^{\dagger}_{i,\alpha}\left(\mathbf{m}_{i,\alpha}\cdot\boldsymbol{\sigma}\right)c_{i,\alpha},
    \end{aligned}
    \label{equ:2D-H}
\end{equation}
where the last term, $H_{sd}$, describes the $s$-$d$ exchange coupling between itinerant electrons and local magnetic moments with coupling strength $J_H$. The remaining terms represent the kinetic energy $H_{kin}$. Here $\langle ij\rangle_x$ and $\langle ij\rangle_y$ denote the nearest-neighbor bonds along $x$ and $y$ directions, and $\langle ij\rangle_{AB}$ is the nearest-neighbor bond connecting sublattice A and B. The operators $c_{i,\alpha,\sigma}$ ($c^{\dagger}_{i,\alpha,\sigma}$) annihilates (creates) electrons with spin $\sigma\in\{\uparrow,\downarrow\}$ on site $i$ of sublattice $\alpha\in\{A,B\}$. $\mathbf{m}_{i,\alpha}$ and $\boldsymbol{\sigma}$ are the local magnetic moments and Pauli matrices. Diagonalizing the Hamiltonian in Eq.~\eqref{equ:2D-H} yields~\cite{Altermagnet_skyrmions:Macroscopic_realization_of_d-wavesymmery}:
\begin{equation}
    E_{\pm}(\sigma)=\epsilon_A+\epsilon_B\pm
    \sqrt{\left(\epsilon_A-\epsilon_B-\sigma mJ_H\right)^2+t_3^{\,2}T^2},
    \label{equ:E}
\end{equation}
with
\begin{equation}
\left\{
\begin{aligned}
\epsilon_A(\mathbf{k}) &= -t_1\cos(k_x a)-t_2\cos(k_y a),\\
\epsilon_B(\mathbf{k}) &= -t_2\cos(k_x a)-t_1\cos(k_y a),\\
T(\mathbf{k}) &= 4\cos\left(k_x a/2\right)
\cos\left(k_y a/2\right).
\end{aligned}
\right.
\end{equation}
In Eq.~\eqref{equ:E}, $\sigma\in\{\uparrow,\downarrow\}=\{+1,-1\}$. Here, we assume that the local magnetic moments ($m$) on different sublattices A and B have the same magnitude. As shown in Fig.~\figsubref{fig:4}{b}, the 2D Fermi contours display the expected $d$-wave AM character, with a spin splitting that changes sign under a $90^\circ$ rotation in reciprocal space. 

Among the four contours in Fig.~\figsubref{fig:4}{b}, we focus on the two closed ones around the BZ center. Their simple closed-pocket geometry makes them suitable analogs to the Fermi pockets of band 5 and 6 in hole-doped \ch{RuO2}. In this 2D model, we do not attempt to compute quantum oscillation frequencies directly. Instead, we characterize the spin splitting of the closed Fermi surface contours by their angle-dependent radii to construct a measure analogous to the quantum-oscillation-based spin splitting. Following the approach used in Eq.~\eqref{equ:Delta} and \eqref{equ:Delta2}, we define the following quantity to describe the spin splitting in this minimal model:
\begin{equation}
    \delta = \frac{\left\langle \left|l_r - l_b\right| \right\rangle}{\left\langle (l_r + l_b)/2 \right\rangle},
    \label{equ:delta}
\end{equation}
where $l_r(\alpha)$ and $l_b(\alpha)$ are the angle-dependent radii of the Fermi contours in red and blue, respectively, and $\left\langle\cdots\right\rangle$ denotes the average over $\alpha\in[-\pi/4,\pi/4]$ in Fig.~\figsubref{fig:4}{b} using 361 sampling points. The convergence with respect to the angular sampling was verified by increasing the number of sampling points to 721, which changes the resulting $\delta$ by less than $0.1\%$. A comparison of the results obtained using 361 and 721 sampling points is presented in Fig.~S3(b) in the SM~\cite{SupplementaryMaterial}. Therefore, Eq.~\eqref{equ:delta} provides a normalized average measure of the radius difference between the two contours. Because SOC only weakly modifies the overall electronic structure of \ch{RuO2}, its omission in the minimal model is justified at a qualitative level. The radius-based quantity defined by Eq.~\eqref{equ:delta} serves physically as the 2D analog of the quantum-oscillation-based signature of spin splitting in \ch{RuO2}, characterizing the spin splitting in this model.

Fig.~\figsubref{fig:4}{c} plots the spin-splitting measure $\delta$ as a function of $mJ_H$ for different parameter sets. One can see that the curves in Fig.~\figsubref{fig:4}{c} are characterized by an initial quasi-linear increase, followed by a gradual saturation as $mJ_H$ increases. Because the quantity $mJ_H/t_1$ is directly proportional to the magnetic moment $m$ for a given system with a fixed value of $J_H/t_1$, $mJ_H$ can be regarded as a measure of $m$ in this model. Therefore, the trend in Fig.~\figsubref{fig:4}{c} reveals that the spin splitting in this AM model is enhanced nearly linearly with increasing $m$, followed by a gradual crossover toward saturation at large $m$. Beyond the three representative parameter sets presented above, additional calculations covering all four relative-magnitude regimes of $t_2$ and $t_3$ with respect to $t_1$ are provided in Fig.~S6 of the SM~\cite{SupplementaryMaterial}. The results further demonstrate that the quasi-linear behavior in the small-moment regime persists over a broad range of model parameters. The behavior in Fig.~\figsubref{fig:4}{c} and Fig.~S6 can be understood from the analytic expression of the spin splitting: 
\begin{equation}
\begin{split}
\Delta E = 
& \Biggl| \sqrt{ \bigl( \epsilon_A - \epsilon_B + m J_H \bigr)^2 + t_3^2\, T^2 } \\
& \quad - \sqrt{ \bigl( \epsilon_A - \epsilon_B - m J_H \bigr)^2 + t_3^2\, T^2 } \Biggr|.
\end{split}
\end{equation}
In the limits $m J_H \!\to\! 0$ and $mJ_H\!\to\!\infty$, the spin splitting $\Delta E$ approaches $2 \, \Delta \epsilon \, m J_H/\sqrt{ (\Delta \epsilon)^2 + t_3^2 \, T^2 }$ and $2\Delta\epsilon$, respectively, where $\Delta\epsilon(\mathbf{k})=\epsilon_A-\epsilon_B$. Upon integration, $\Delta E$ shows a linear dependence on $mJ_H$ in the small magnetic moment limit, and saturates to a constant in the large magnetic moment limit. This behavior agrees well with the numerical result in Fig.~\figsubref{fig:4}{c}, and also with the general trend reported previously~\cite{trend1,trend2}, supporting Eq.~\eqref{equ:Delta} and \eqref{equ:delta} as spin splitting measures. Overall, these results indicate that, within the 2D $d$-wave AM minimal model, the spin splitting increases monotonically with magnetic moment and remains approximately linear for the small magnetic moment.

In the minimal 2D model and DFT calculated hole-doped \ch{RuO2}, the spin splitting shows a monotonic, quasi-linear dependence on the magnetic moment in the small magnetic moment region. This agreement between these two results shows that the minimal 2D model intuitively captures the small magnetic moment behavior in the hole-doped \ch{RuO2}, namely that the AM spin splitting is enhanced quasi-linearly as the magnetic moment increases. This further suggests that such a correlation may point to a broader feature of $d$-wave altermagnets rather than an isolated case, given that this trend already appears in a 2D minimal model only with essential symmetry ingredients. However, unlike the minimal model, the results of the doped \ch{RuO2} do not show a clear saturation plateau. This difference is likely related to the fact that in-plane anisotropy in \ch{RuO2} primarily originates from next-nearest-neighbor hopping~\cite{minimal_model}. However, the minimal 2D model in Fig.~\figsubref{fig:4}{a} is restricted to nearest-neighbor hopping. To address this, we further consider a more realistic model for hole-doped \ch{RuO2} in Section S3 in the SM~\cite{SupplementaryMaterial}. In this case, the agreement between this 2D model and DFT calculated hole-doped \ch{RuO2} results is greatly improved. Overall, the consistency between the minimal 2D model predictions and DFT calculations suggests that the quasi-linear dependence of spin splitting on magnetic moments may provide guidance for identifying the AM state in the hole-doped \ch{RuO2} through quantum oscillations.

From the perspective of experimental probes, it may be challenging to directly identify the quantum oscillation spectra of band 5 and 6 from the full set of complex signals. However, as discussed above, the two spectral branches of band 5 and 6 exhibit a strong angular dependence. These two branches are degenerate at $\theta=0$ and separate at finite $\theta$, consistent with the underlying $d$-wave symmetry. In addition, the simple closed-pocket geometry of their Fermi surfaces ensures the continuity of their spectra. Such characteristics provide effective criteria for identifying the two frequency branches in the low-frequency window, which should be considered simultaneously for reliable identification in experiments. The quantum oscillation spectra within the low-frequency window from 3 to $9\,\mathrm{kT}$ are shown in Fig.~S2 in the SM~\cite{SupplementaryMaterial}, where the branches of band 5 and 6 are identifiable. Therefore, for future relevant experimental studies, attention should be mainly paid to the low-frequency region, with sufficiently fine sampling employed to obtain relatively continuous spectra.

\begin{figure*}[t]
    \centering
    \includegraphics[width=1\linewidth]{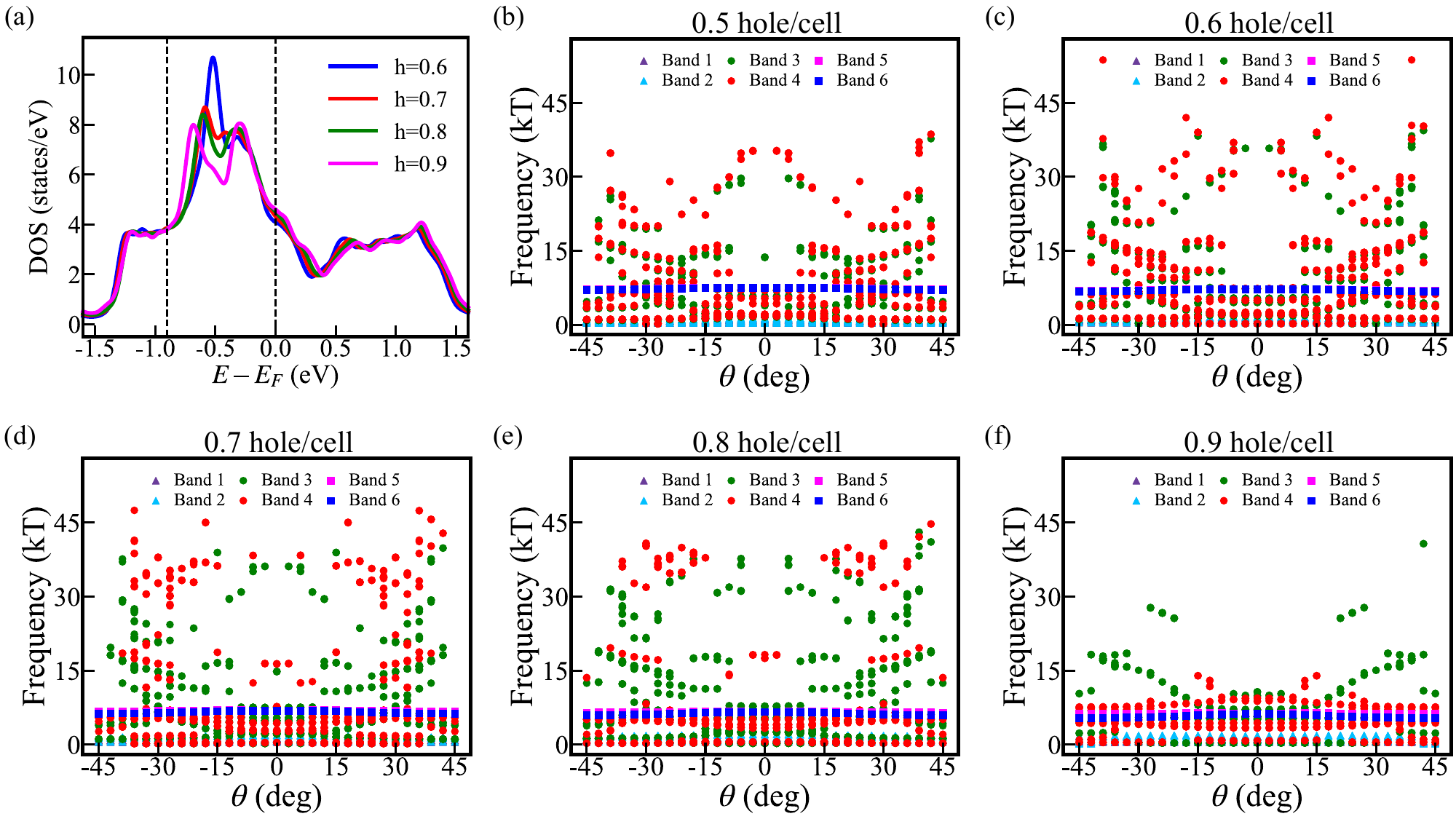}
    \caption{Total DOSs and quantum oscillation frequencies of the intermediate state in hole-doped \ch{RuO2}. (a) DOS of hole-doped \ch{RuO2} with hole doping from 0.6 to 0.9~hole/cell. (b)-(f) Quantum oscillation frequency spectrum of hole-doped \ch{RuO2} with hole doping from 0.5 to 0.9~hole/cell.}
    \label{fig:5}
\end{figure*}

\subsection{Intermediate state in the NM-AM transition of hole-doped \ch{RuO2}}

Fig.~\figsubref{fig:3}{b} also shows two noticeable jumps in both the local magnetic moment of Ru and $\Delta$ in the white-shaded region. The first jump emerges within the range of 0.6--0.7~hole/cell. Below 0.7~hole/cell, both $m_{Ru}$ and $\Delta$ remain relatively low, aside from minor numerical fluctuations. At such hole-doping levels, \ch{RuO2} can be considered to be in a nearly NM state, with almost no magnetic moment in real space and negligible spin splitting in reciprocal space. In this region, the narrow DOS peak near the Fermi energy is still present. Once the doping exceeds about 0.7~hole/cell, both $m_{Ru}$ and $\Delta$ simultaneously exhibit an abrupt increase to a notable magnitude, from $0.04\,\mu_B$ and 0.0122 to $0.138\,\mu_B$ and 0.0411, respectively. The simultaneous jump in magnetic moment and spin splitting indicates that hole-doped \ch{RuO2} undergoes a magnetic transition from an NM state to an AM state around 0.7~hole/cell. The second marked change appears within 0.8--0.9~hole/cell. At this point, both quantities jump again, from $0.17\,\mu_B$ and 0.0507 to $0.345\,\mu_B$ and 0.0934, respectively. To further clarify this anomaly, we examine the DOS of \ch{RuO2} in the doping range spanning the two jumps and show the results in Fig.~\figsubref{fig:5}{a}. We see that the DOS remains nearly identical outside the energy window approximately from $-0.9\,\mathrm{eV}$ to $0.0\,\mathrm{eV}$. Near the Fermi level, however, the DOS exhibits two abrupt changes at 0.7 and 0.9~hole/cell, matching the two jumps seen in Fig.~\figsubref{fig:3}{b}. For the first change in the DOS at 0.7~hole/cell, the narrow peak around $-0.5\,\mathrm{eV}$ drops sharply. Within a modified Stoner picture, this reduction is consistent with an increased tendency toward magnetic stabilization, as reported previously~\cite{Fragility_of_the_magnetic_order}. When the doping level exceeds about 0.9~hole/cell, the remaining lower peak is further reduced and separates into two distinct peaks. In addition, as shown in Fig.~S8 in the SM~\cite{SupplementaryMaterial}, band 4 develops an additional Fermi-level crossing along a certain high-symmetry path near 0.9~hole/cell, consistent with the DOS evolution. The abrupt changes in Fig.~\figsubref{fig:3}{b}, Fig.~\figsubref{fig:5}{a} and Fig.~S8 support the same picture, suggesting the presence of an intermediate state emerging between the NM and AM states in the hole-doped \ch{RuO2}.

The quantum oscillation frequency spectra of intermediate hole doping levels further support the intermediate state in hole-doped \ch{RuO2}. As shown in Fig.~\figsubref{fig:5}{b}--\figsubref{fig:5}{e}, the spectra are basically similar and remain highly complex for doping levels from 0.5 to 0.8~hole/cell, although the magnetic transition appears within the range of 0.6--0.7~hole/cell. The complexity of the spectrum reflects a highly perforated Fermi-surface topology, in which well-defined closed pockets are difficult to form, particularly for band 3 and 4. By contrast, a sudden change of the quantum oscillation spectrum occurs within the range of 0.8--0.9~hole/cell. In Fig.~\figsubref{fig:5}{f}, the spectrum is strongly simplified and becomes clearly distinguishable, with high-frequency components above $30\,\mathrm{kT}$ almost completely disappearing. This change reveals a significant reconstruction of the Fermi surface topology in hole-doped \ch{RuO2}, with large perforated sheets changing to smaller, well-defined closed pockets. Meanwhile, the frequencies in Fig.~\figsubref{fig:5}{f} are also distinct from those of the NM state in Fig.~\figsubref{fig:2}{c}: although the quantum oscillation spectra have similar overall scale, their distribution is smoother, more continuous and more compact. Hence, the distinct quantum oscillation frequency in the high-doping AM regime may serve as a useful signature of the AM state in hole-doped \ch{RuO2}.

The two abrupt changes observed near 0.7 and 0.9~hole/cell may reflect different aspects of the NM–AM transition. The first change is accompanied by a simultaneous increase in $m_{Ru}$ and $\Delta$, suggesting the onset of magnetic instability and notable spin splitting. The second change is primarily manifested by a substantial simplification of the quantum oscillation spectrum, the splitting of the DOS peak into two distinct peaks, and the emergence of an additional Fermi-level crossing of band 4. These behaviors point to a pronounced change in the electronic structure. Taking the two abrupt changes together, we thereby  refer to the range between these two doping levels as an intermediate regime.

\section{Conclusion}

In summary, we have performed first-principles calculations to study quantum oscillations in hole-doped \ch{RuO2}. We show that hole doping enhances spin splitting and drives a systematic evolution of the magnetic and electronic structure, which can be tracked through quantum oscillations. A robust quasi-linear correlation is revealed between the local magnetic moment of Ru and a quantum-oscillation-based signature of AM spin splitting. This trend can be understood using a minimal 2D $d$-wave AM model, particularly in the small magnetic moment region, suggesting that similar behavior may occur in $d$-wave altermagnets. Our results further support the presence of an intermediate state in the NM-AM transition in hole-doped \ch{RuO2}. The characteristic quantum oscillation frequency of the stable AM state, together with the robust correlation, provides clear experimental signatures. These findings offer an understanding on the interplay between magnetic moment and spin splitting in hole-doped \ch{RuO2} and provide guidance for identifying AM states in future experiments.

\section*{ACKNOWLEDGMENTS}

This work was supported by the National Key R\&D Program of China (Grant No.~2024YFA1408303, 2022YFA1403301), the National Natural Sciences Foundation of China (Grants No.~12474247 and 92165204). Yusheng Hou acknowledges the support from Guangdong Provincial Key Laboratory of Magnetoelectric Physics and Devices (Grant No.~2022B1212010008) and Research Center for Magnetoelectric Physics of Guangdong Province (Grant No.~2024B0303390001). Density functional theory calculations are performed on Tianhe-Xingyi.

\section*{DATA AVAILABILITY}

The data that support the findings of this study are available from the authors upon reasonable request.

\FloatBarrier
\bibliography{reference}

\end{document}


\preprint{APS/123-QED}

\title{SUPPLEMENTARY MATERIAL:\\Quantum oscillation fingerprints of altermagnetism in hole-doped \ch{RuO2}}

\author{Yuchi Yang}

\author{Yusheng Hou}
\email{houysh@mail.sysu.edu.cn}

\affiliation{
Guangdong Provincial Key Laboratory of Magnetoelectric Physics and Devices, Institute of Neutron Science and Technology, School of Physics, 
Sun Yat-Sen University, Guangzhou 510275, China
}

\maketitle


\renewcommand{\theequation}{S\arabic{equation}} 
\renewcommand{\thefigure}{S\arabic{figure}}

\renewcommand{\thesection}{SECTION S\arabic{section}}
\newcommand{\figsubref}[2]{\hyperref[#1]{\ref*{#1}(#2)}}

\section{The evolution of quantum oscillation frequency}

\begin{figure}[H]
    \centering
    \includegraphics[width=\linewidth]{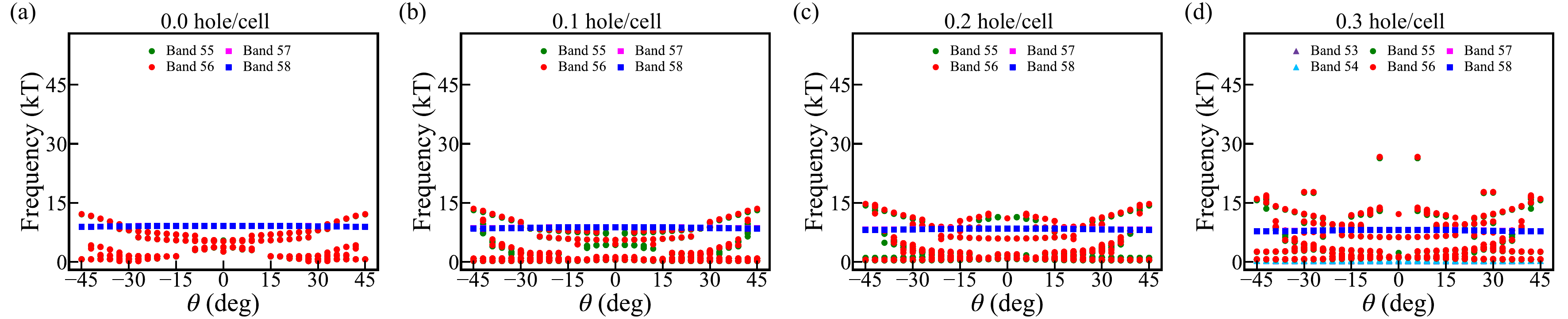}
    \includegraphics[width=\linewidth]{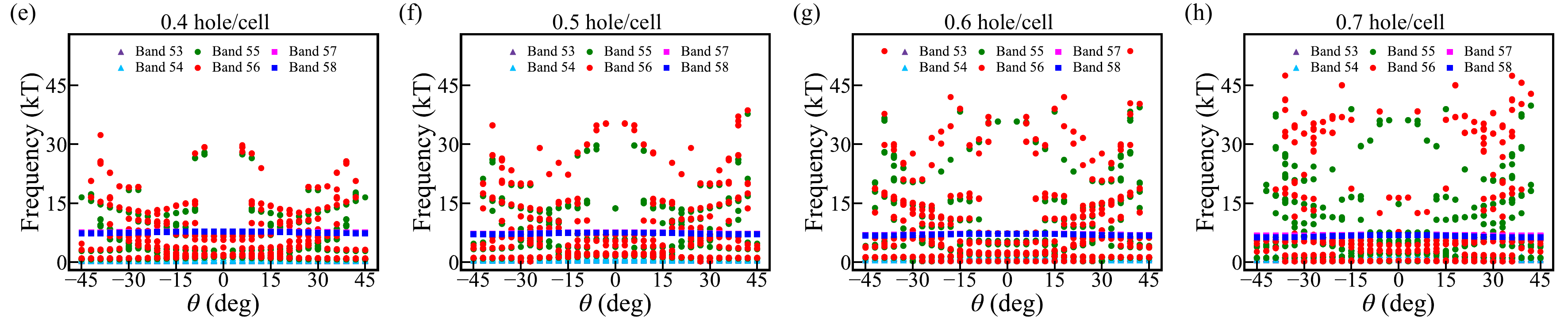}
    \includegraphics[width=\linewidth]{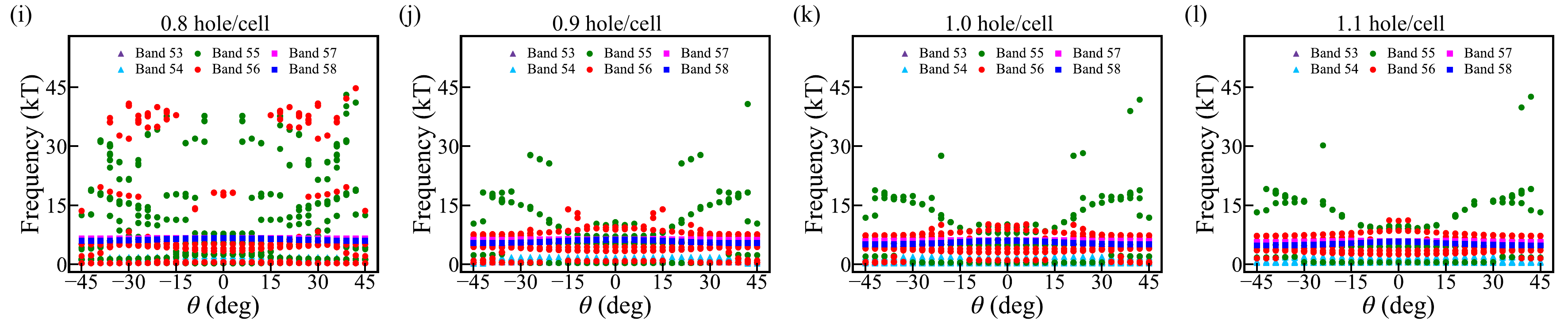}
    \includegraphics[width=\linewidth]{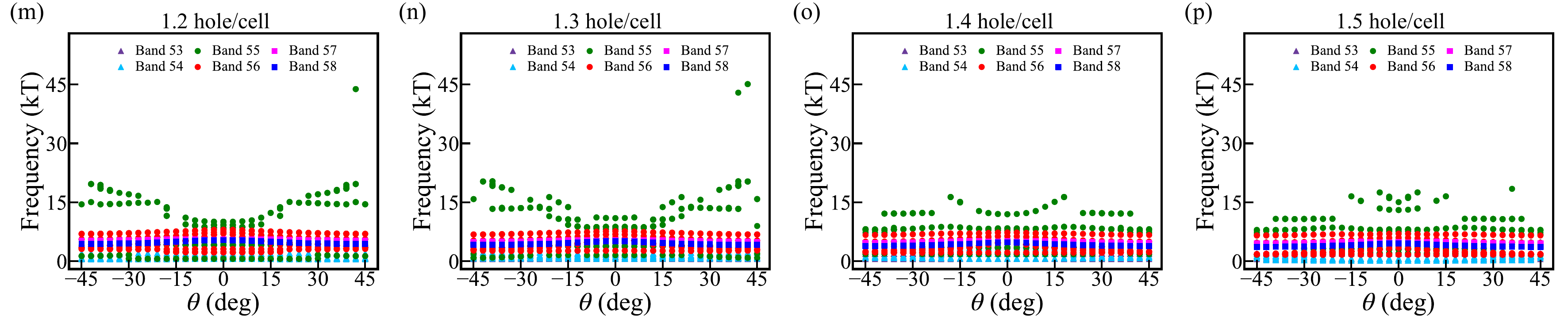}
    \caption{The angle-dependent quantum oscillation frequencies of \ch{RuO2} with doping levels from 0.0 to 1.5~hole/cell in steps of 0.1 per cell.}
    \label{fig:SM1}
\end{figure}

Fig.~\ref{fig:SM1} shows the evolution of the angle-dependent quantum oscillation frequencies in \ch{RuO2} under hole doping, reflecting the gradually enhanced spin splitting. With doping levels below 0.9~hole/cell, the spectrum generally becomes complicated, with data points becoming increasingly scattered, particularly in the high-frequency region above $15\,\mathrm{kT}$. Besides, an overall increase in frequency is observed. Within the range of 0.6--0.7~hole/cell, the change of the spectrum remains relatively modest although the magnetic transition emerges. For doping levels above 0.9~hole/cell, however, the spectrum suddenly becomes much simpler and thus distinct, with most high-frequency branches exceeding $20\,\mathrm{kT}$ disappearing. Above this doping level, the spectrum no longer changes sharply. The spectrum becomes progressively simpler, with the frequency branches forming a smoother and more continuous distribution. Meanwhile, the maximum frequency decreases slightly. This evolution of the quantum oscillation frequency supports the division into the NM state, intermediate and AM state regime discussed in the main text.

\begin{figure}[H]
    \centering
    \includegraphics[width=\linewidth]{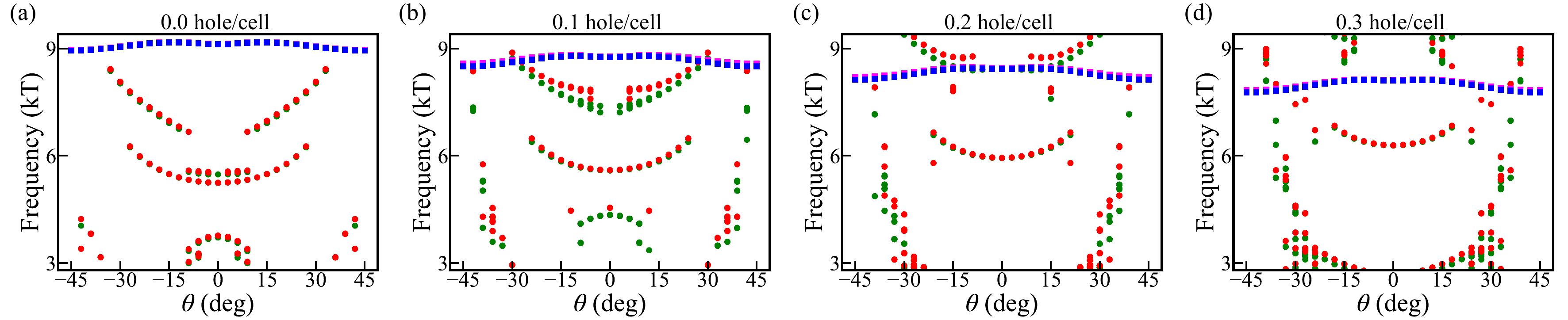}
    \includegraphics[width=\linewidth]{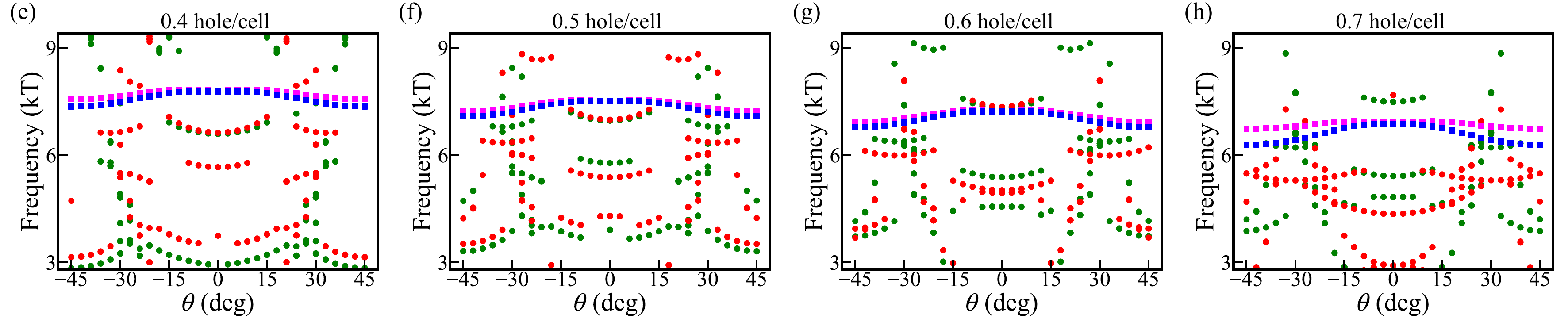}
    \includegraphics[width=\linewidth]{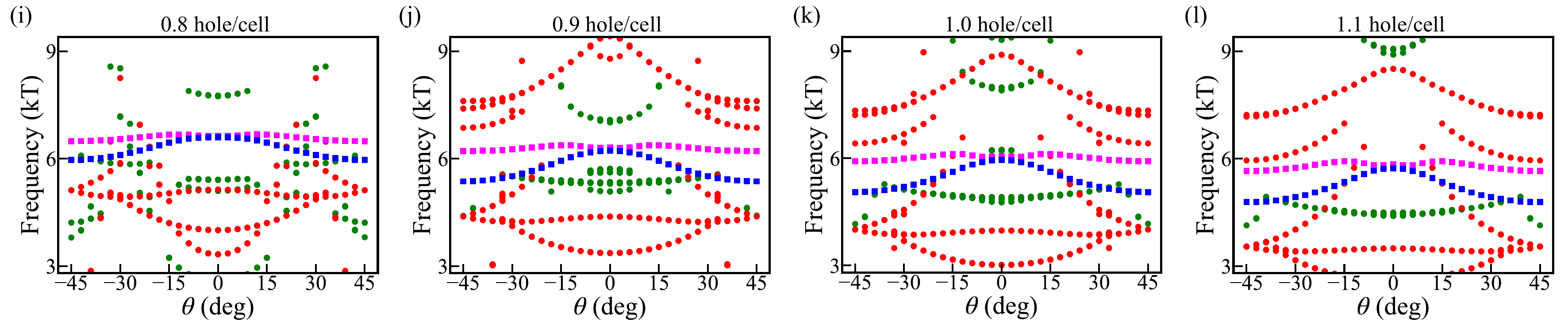}
    \includegraphics[width=\linewidth]{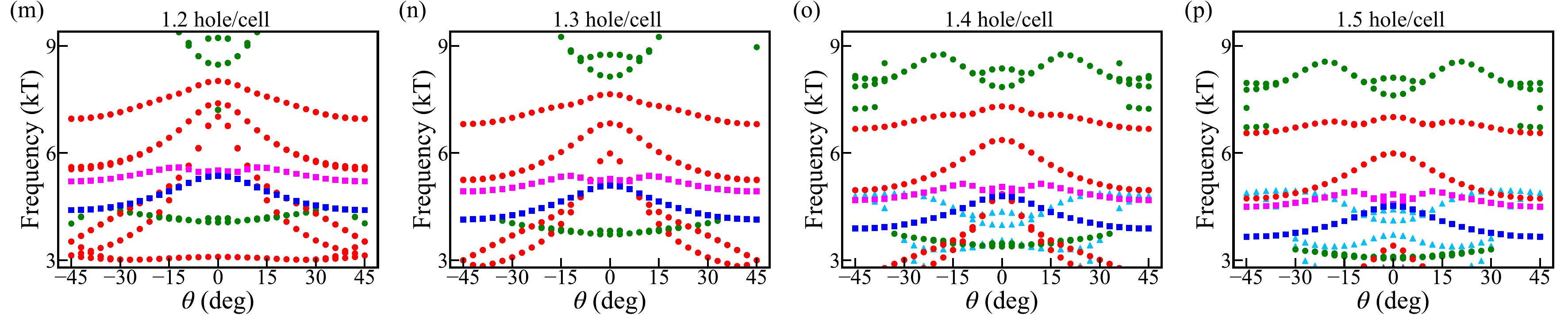}
    \caption{The angle-dependent quantum oscillation frequencies in the low-frequency window from 3 to 9~kT of \ch{RuO2} with doping levels from 0.0 to 1.5~hole/cell in steps of 0.1 per cell. The labels are the same as Fig. S1. The purple and blue square symbols correspond to the spectra of band 5 and 6 that need to be identified.}
    \label{fig:SM1-low}
\end{figure}

Fig.~\ref{fig:SM1-low} shows the evolution of the angle-dependent quantum oscillation frequencies in the low-frequency window from 3 to $9\,\mathrm{kT}$. When spin splitting is negligible for doping levels below 0.6~hole/cell, only two overlapping sets of frequency spectra associated with the band 5 and 6 form continuous angular branches. As the spin splitting becomes enhanced in the case of doping above 0.6~hole/cell, additional continuous branches emerge. However, only two branches are degenerate at $\theta=0$ and separate at finite $\theta$, consistent with the $d$-wave symmetry. Therefore, the branches of band 5 and 6 can be identified throughout the doping evolution, by considering these characteristic features together with sufficiently dense angular sampling in future experimental probes.

\section{The evolution of magnetic moment and normalized relative difference in different cases}

In this part, we calculate and show the evolution of the local Ru magnetic moment and the normalized relative difference $\Delta$ with hole doping under different conditions, as shown in Fig.~\ref{fig:additional-two-curves}. These cases include (i) different on-site Hubbard values ($U=0.2\,\mathrm{eV}$ and $U=0.6\,\mathrm{eV}$), (ii) hydrostatic strain on the entire lattice, and (iii) in-plane epitaxial strain with the out-of-plane $c$ axis fixed. The lattice constants are scaled to $99\%$ or $101\%$. In all the cases, the two curves evolve in a similar manner, sharing the same abrupt change point, while the transition point from NM to AM exhibits negligible change. In the results of $U=0.2\,\mathrm{eV}$, the consistency of the two curves is great, while the two curves of $U=0.6\,\mathrm{eV}$ are less consistent. The less consistency in $U=0.6\,\mathrm{eV}$ may be attributed to the relatively large Hubbard value for \ch{RuO2}. In the strain cases, the magnetic moment is generally lower under compressive strain than under tensile strain, as reported before~\cite{Fragile2}. However, the magnitude of $\Delta$ is slightly larger under tensile strain in isotropic cases, while the magnitude is larger under compressive strain in in-plane cases. Overall, the results are intuitively similar to the result in Fig.~3(b), supporting the robustness.

Fig.~\ref{fig:additional-m-to-splitting} further shows the dependence of the quantum-oscillation-based signature of spin splitting on the local Ru magnetic moment $m_{Ru}$ under different strain conditions. The results for different isotropic strain cases follow a similar trend, although the in-plane strain results deviate more noticeably from the unstrained structure. Nevertheless, all strained cases exhibit quasi-linear behavior similar to that in Fig.~3(c), supporting the robustness of the quasi-linear correlation discussed in the main text.

\section{Beyond the Minimal Model: Robustness Analysis and A More Realistic 2D Model}

In the main text, we discuss a minimal 2D tight-binding model to qualitatively capture the quasi-linear correlation observed in \ch{RuO2}. Although three representative parameter sets have been presented in the main text, a more systematic exploration of the parameter space is useful to further verify the robustness of this behavior. Fig.~\ref{fig:fourspace} shows the results of different parameter sets covering all four relative-magnitude regimes defined by whether $t_2$ and $t_3$ are smaller or larger than $t_1$, namely, ($t_2,t_3<t_1$), ($t_3<t_1<t_2$), ($t_2<t_1<t_3$) and ($t_2,t_3>t_1$). Across all tested cases, variations in the model parameters modify both the slope of the quasi-linear regime and the onset of saturation, but the quasi-linear correlation in the small-moment regime remains clearly identifiable. These results indicate that the quasi-linear behavior persists over the explored parameter range.

Beyond the minimal model, we further construct a more realistic model for \ch{RuO2}, which is extended from the original minimal model with additional corrections. In \ch{RuO2}, the in-plane anisotropy primarily originates from next-nearest-neighbor hopping rather than nearest-neighbor hopping~\cite{minimal_model}. As shown in Fig.~\figsubref{fig:realistic-model}{a}, the in-plane nearest-neighbor hopping $t_1$ and $t_2$ are set equal to $t$, while $t_4$ and $t_5$ denote the in-plane next-nearest-neighbor hopping. Bonds with the same color in Fig.~\figsubref{fig:realistic-model}{a} are assigned the same hopping amplitude. With these modifications, Eq.~(5) retains the same functional form, with only the definitions of $\epsilon_A$ and $\epsilon_B$ modified as:
\begin{equation}
    \left\{
    \begin{aligned}
        \epsilon_A(\mathbf{k}) &= -\Big[
        t\big(\cos(k_x a)+\cos(k_y a)\big)
        + t_4 \cos(k_x a+k_y a)
        + t_5 \cos(k_x a-k_y a)
        \Big], \\
        \epsilon_B(\mathbf{k}) &= -\Big[
        t\big(\cos(k_x a)+\cos(k_y a)\big)
        + t_5 \cos(k_x a+k_y a)
        + t_4 \cos(k_x a-k_y a)
        \Big].
    \end{aligned}
    \right.
    \label{S1}
\end{equation}

In addition, we further account for the reduction of the Fermi energy induced by hole doping. In hole-doped \ch{RuO2}, a linear fit yields $E_f=7.94-0.55h$, where $E_f$ is the Fermi energy in eV, and $h$ denotes the hole doping level in units of hole/cell. Since increasing hole doping is accompanied by an enhancement of the local moment, we phenomenologically assume that the Fermi energy decreases linearly with $mJ_H$. Specifically, $E_f=E_{f0}-\beta(mJ_H)$. For the hopping parameters, we adopt the parameter set $t=0.05$, $t_3=0.425$, $t_4=-0.1$, $t_5=0.05$, as proposed in Ref.~\cite{minimal_model}. Fig.~\figsubref{fig:realistic-model}{b} presents the results obtained using this reference parameter set, together with several nearby parameter sets generated by small variations around it, with $a=1$ and $E_f=-0.5-mJ_H$. This choice of $E_{f0}=-0.5$ and $\beta=1$ ensures that the split bands continue to cross the Fermi energy. The results all exhibit a quasi-linear trend similar to that in Fig.~3c, without the saturation plateau observed in the nearest-neighbor model. Moreover, the value $J_H=0.2$ is adopted, following Ref.~\cite{minimal_model}. Therefore, the quasi-linear regime extends over a relatively broad range of local moment $m$, further supporting the quasi-linear correlation in hole-doped \ch{RuO2}.

\section{Additional discussion about the intermediate state}

Beyond the discussion about the abrupt changes of the DOS and the quantum oscillation frequency spectra presented in the main text, we here provide additional insight into the intermediate state from the perspective of band evolution. As shown in Fig.~\ref{fig:SMband}, for hole doping below 0.9~hole/cell, the relevant segment of band~4 remains below the Fermi level along the high-symmetry path $\Gamma\text{--}X\text{--}M\text{--}\Gamma\text{--}M_1$. When the doping reaches 0.9~hole/cell, this segment crosses the Fermi level, producing an additional Fermi-level crossing of band~4. The newly crossing segment is predominantly derived from Ru-$d$ orbitals. The behavior of band 4 is consistent with the abrupt change in the DOS in the main text, suggesting a pronounced change in the electronic structure near 0.9~hole/cell.

\newpage

\begin{figure}[H]
    \centering
    \includegraphics[width=1\linewidth]{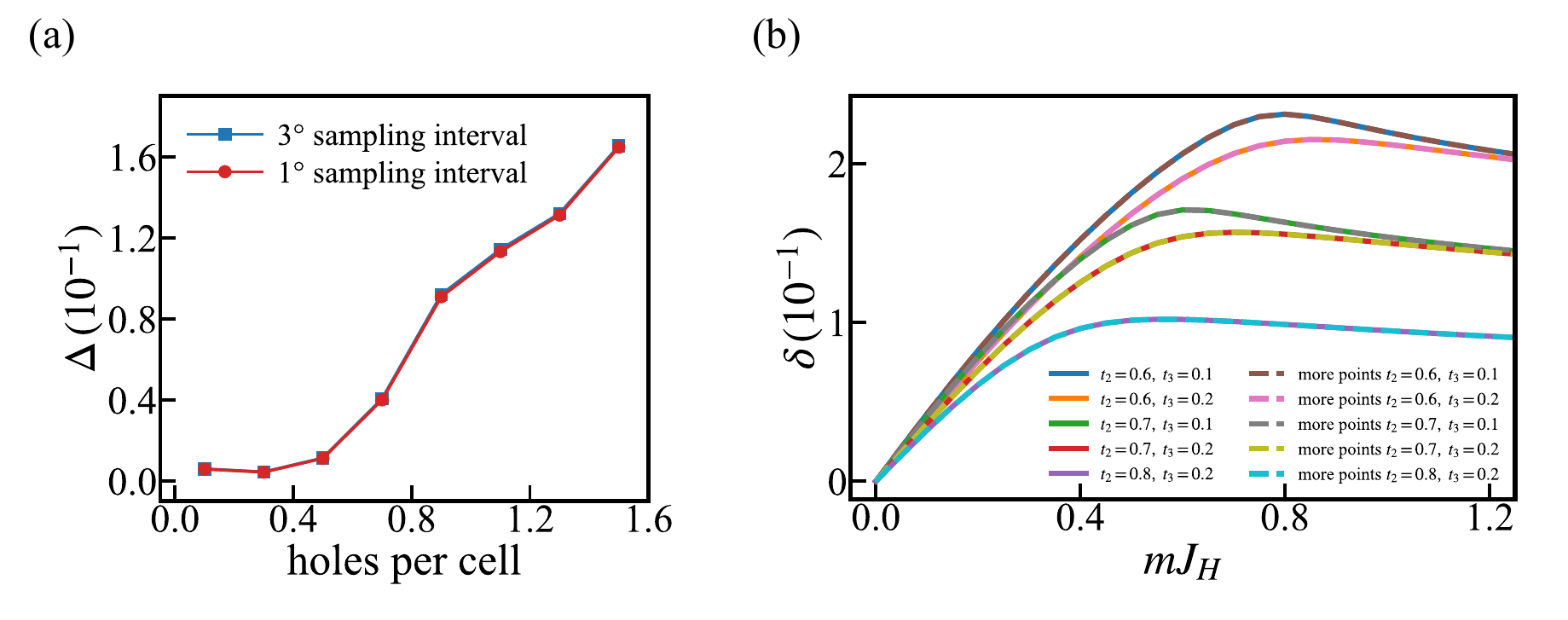}
    \caption{Convergence tests with respect to angular sampling. (a) Hole-doping dependence of $\Delta$ calculated using $3^\circ$ and $1^\circ$ sampling intervals. (b) Dependence of $\delta$ on $mJ_H$ calculated using 361 and 721 sampling points. In panel (b), the solid and dashed lines denote the results obtained using 361 and 721 sampling points, respectively. Among the five parameter sets, the first three listed in the legend correspond to those used in the main text. The nearly overlapping results confirm the convergence of the angular sampling used in the main text.}
    \label{fig:sampling}
\end{figure}

\begin{figure}[H]
    \centering
    \includegraphics[width=\linewidth]{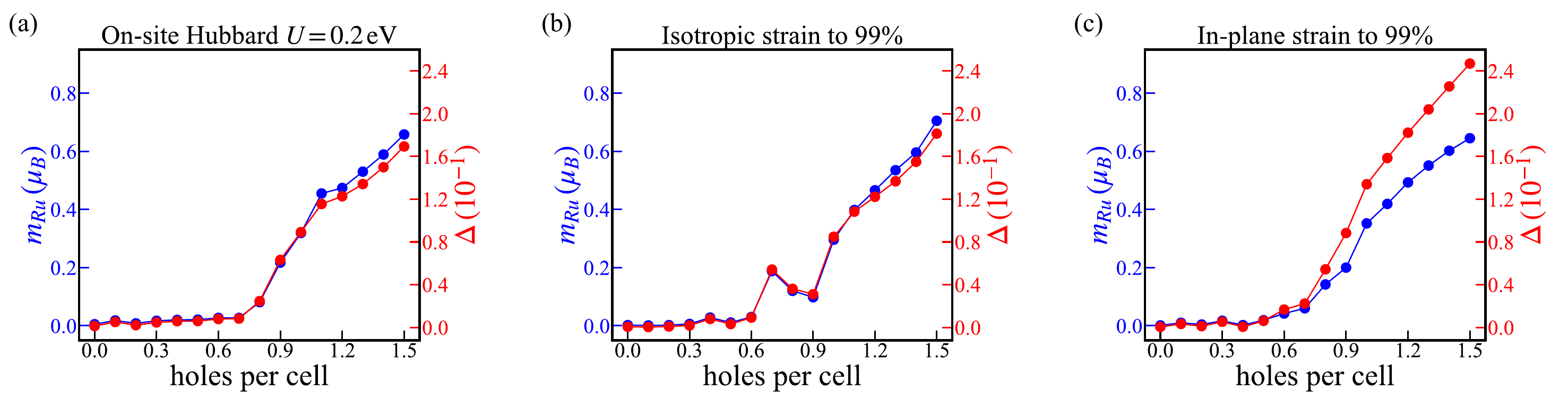}
    \includegraphics[width=\linewidth]{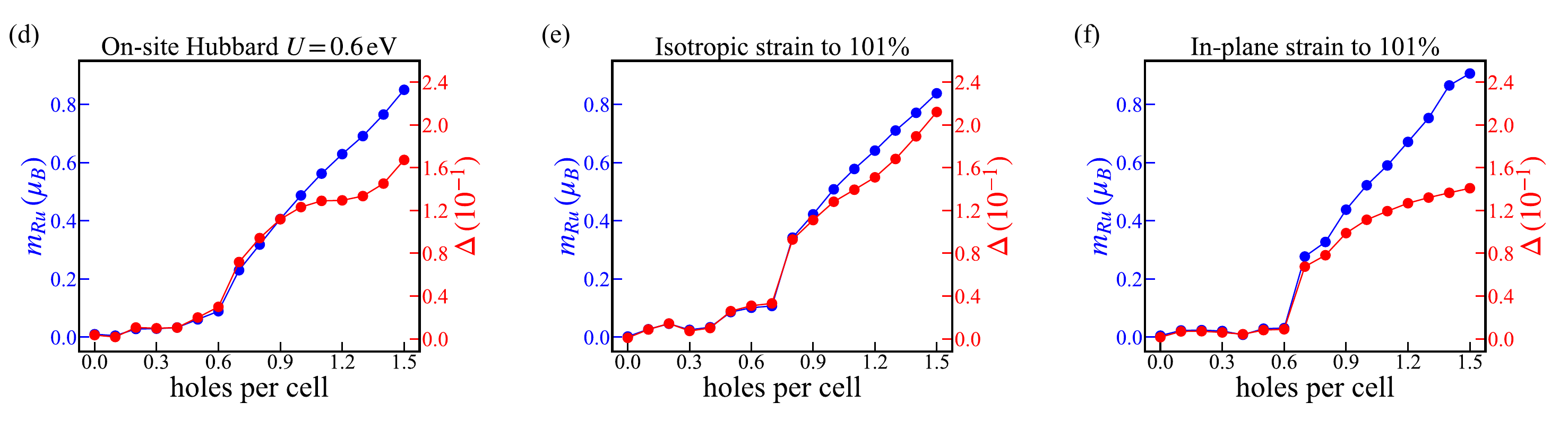}
    \caption{Hole-doping dependence of magnetic moment $m_{Ru}$ (left axis) and the normalized relative difference (right axis) under different conditions. (a), (d) Results for different on-site Hubbard values, $U=0.2\,\mathrm{eV}$ and $U=0.6\,\mathrm{eV}$, respectively. (b), (e) Results for isotropic lattice scaling under tensile and compressive strain. (c), (f) Results for in-plane epitaxial tensile and compressive strain with the out-of-plane c axis fixed. In all strain cases, the tensile and compressive strains are $\pm1\%$}.
    \label{fig:additional-two-curves}
\end{figure}

\begin{figure}[H]
    \centering
    \includegraphics[width=\linewidth]{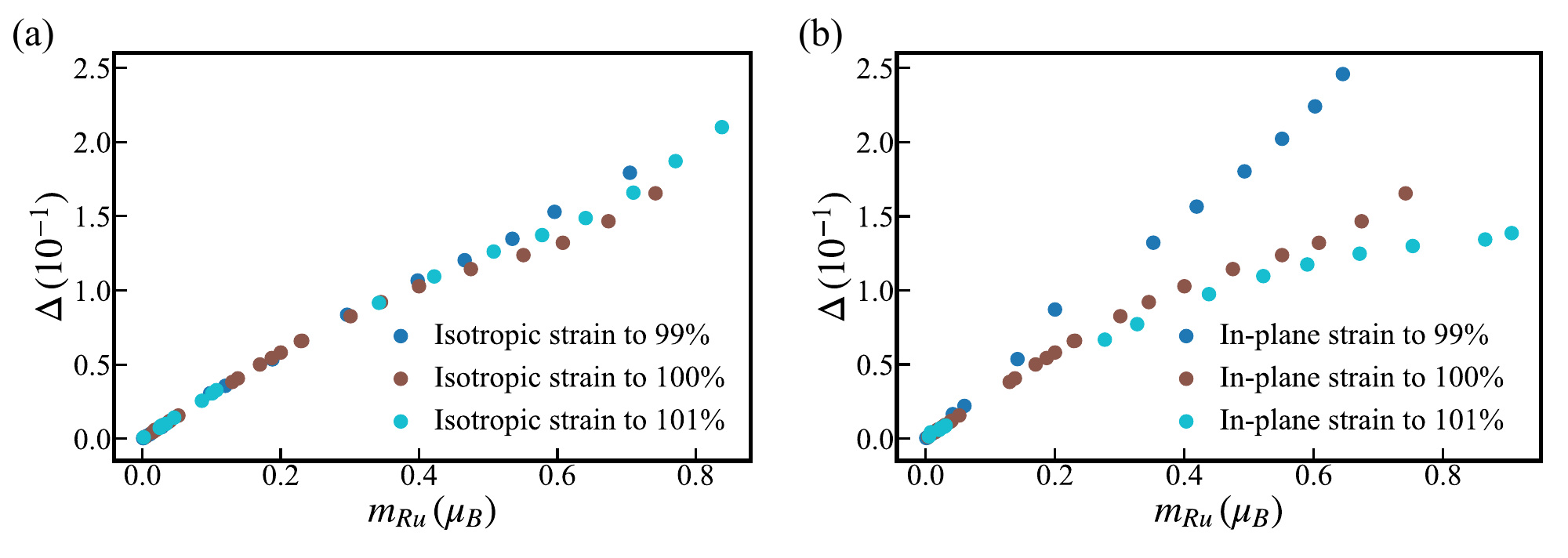}
    \caption{Dependence of $\Delta$ on the magnetic moment $m_{Ru}$ in hole-doped \ch{RuO2} under different strain conditions. (a) Results for hydrostatic strain. (b) Results for in-plane epitaxial strain with the out-of-plane c axis fixed.}
    \label{fig:additional-m-to-splitting}
\end{figure}

\begin{figure}[H]
    \centering
    \includegraphics[width=\linewidth]{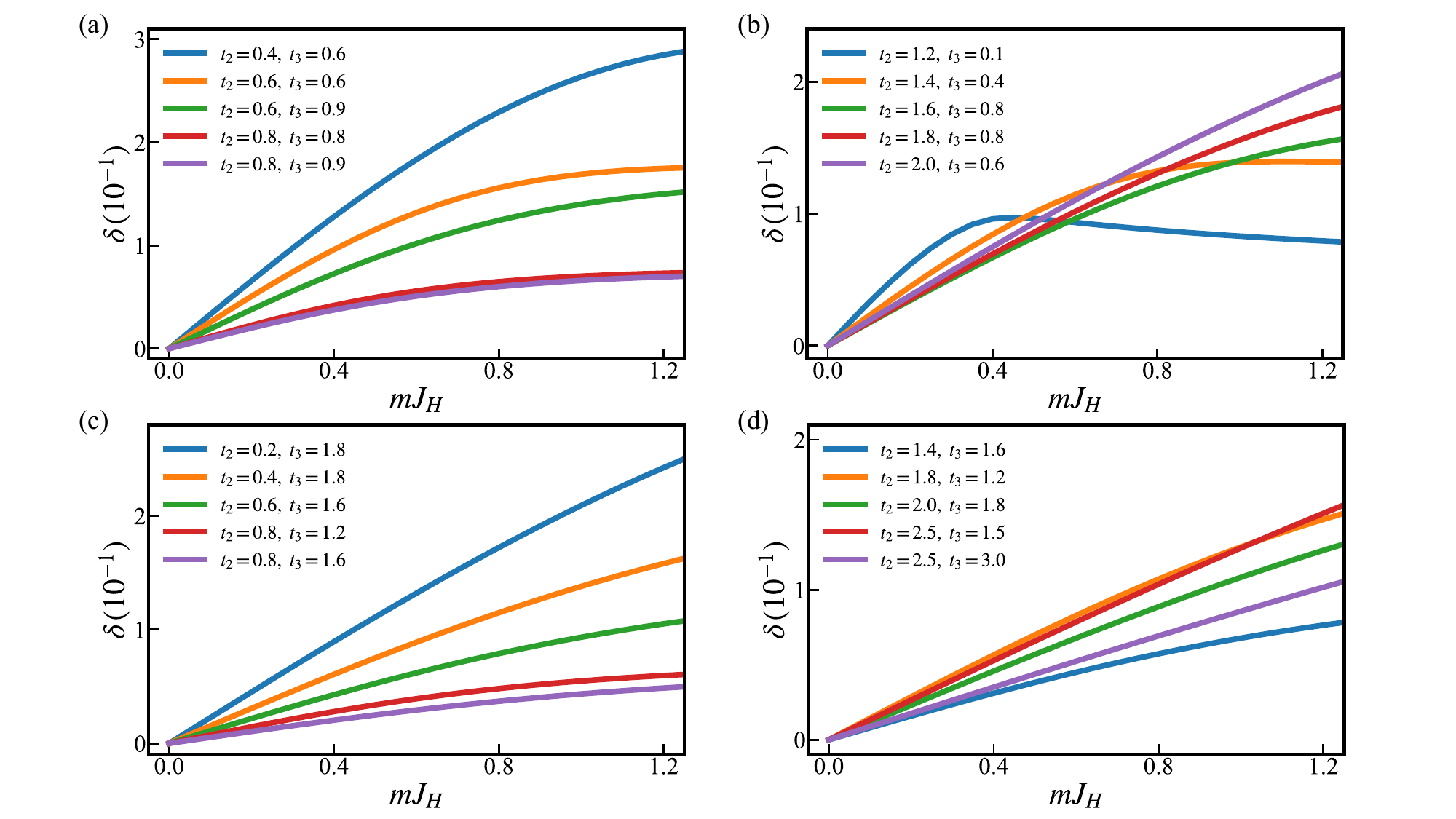}
    \caption{Dependence of $\delta$ on $mJ_H$ for additional parameter sets. The quantities $t_2$, $t_3$, and $J_H$ are in units of $t_1$. (a)-(d) Results for $(t_2,t_3<t_1)$, $(t_3<t_1<t_2)$, $(t_2<t_1<t_3)$ and $(t_2,t_3>t_1)$.}
    \label{fig:fourspace}
\end{figure}

\begin{figure}[H]
    \centering
    \includegraphics[width=\linewidth]{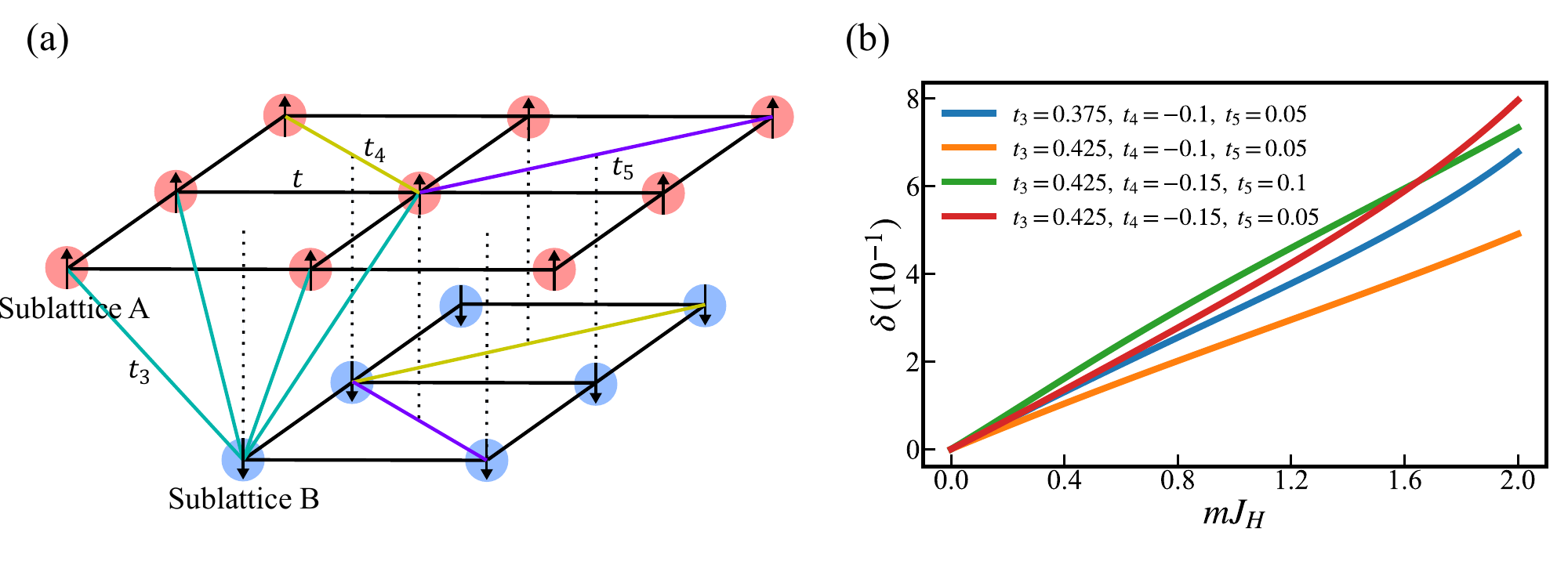}
    \caption{More realistic 2D tight-binding model for AM \ch{RuO2}. (a) Schematic of the 2D lattice. (b) Dependence of $\delta$ on $mJ_H$ using different parameter sets. The orange line corresponds to the result using the reference parameter set.}
    \label{fig:realistic-model}
\end{figure}

\begin{figure}[H]
    \centering
    \includegraphics[width=1.0\linewidth]{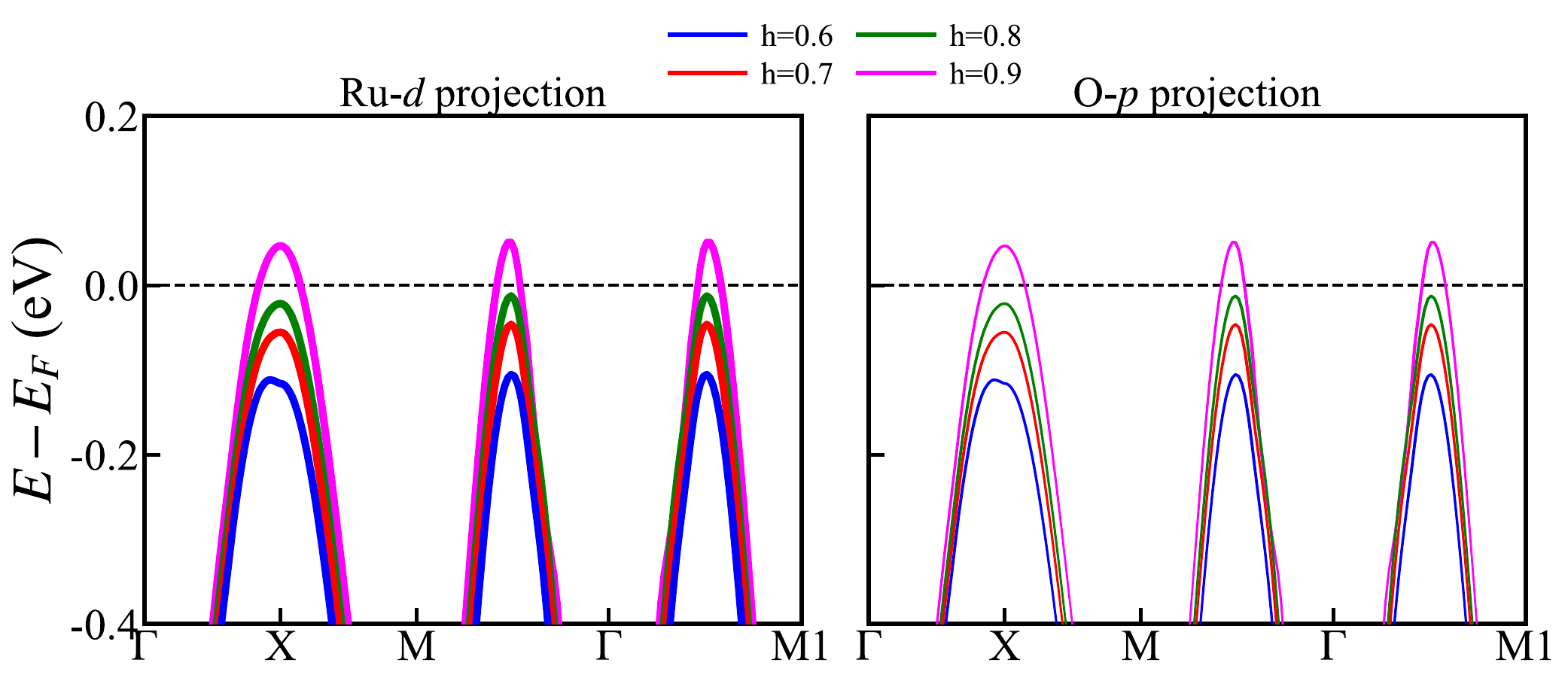}
    \caption{Band 4 of hole-doped \ch{RuO2} in the intermediate regime with hole doping from 0.6 to 0.9~hole/cell. The left and right panels exhibit the Ru-$d$ and O-$p$ orbital projections.}
    \label{fig:SMband}
\end{figure}

\bibliography{reference}